\documentclass[12pt, amssymb,amsmath,aps,showpacs,floatfix,nofootinbib,showpacs,balancelastpage]{revtex4}
\usepackage{hangcaption,fancybox,feynmp}
\usepackage{indentfirst}
\usepackage{graphicx}
\usepackage{epsfig,subfigure,psfrag}
\usepackage{mathrsfs}
\usepackage{color}

\begin{document}

\include{newcmd}
%%%%%%%%%%%%%%%%%%%%%%%%%%%%%%%%%%%

\title{ Edge Charge Asymmetry in Top Pair Production at the LHC }
\author{Bo Xiao$^{1}$\footnote{E-mail:homenature@pku.edu.cn},
You-Kai Wang $^{1}$\footnote{E-mail:wangyk@pku.edu.cn}, Zhong-Qiu
Zhou $^{1}$\footnote{E-mail:zhongqiu\_zhou@pku.edu.cn}, and Shou-hua
Zhu$^{1,2}$\footnote{E-mail:shzhu@pku.edu.cn}}

\affiliation{ $ ^1$ Institute of Theoretical Physics $\&$ State Key
Laboratory of Nuclear Physics and Technology, Peking University,
Beijing 100871, China \\
$ ^2$ Center for High Energy Physics, Peking University, Beijing
100871, China }

\date{\today}
\maketitle

\begin{center}
{\bf Abstract}

\begin{minipage}{15cm}
{ \small
\hskip 0.25cm

In this brief report, we propose a new definition of charge
asymmetry in top pair production at the LHC, namely the edge charge
asymmetry (ECA). ECA utilizes the information of drifting direction
only for single top (or anti-top) with hadronically decay. Therefore
ECA can be free from the uncertainty arising from the missing
neutrino in the $t\bar{t}$ event reconstruction. Moreover rapidity
$Y$ of top (or anti-top) is required to be greater than a critical
value $Y_{\rm C}$ in order to suppress the symmetric $t\bar{t}$
events mainly due to the gluon-gluon fusion process.  In this paper
ECA is calculated up to next-to-leading order QCD in the standard
model and the choice of the optimal $Y_{\rm C}$ is investigated. }

\end{minipage}
\end{center}

%\renewcommand{\baselinestretch}{1.2}
%\fontsize{12pt}{12pt}\selectfont

\newpage

\section{Introduction \label{introduction}}

Being the heaviest fermion ever known, the top quark has many unique
features and it is thought to be closely related with the new
physics beyond the standard model (BSM). After top quark was
discovered in 1994, the measurement of its angular distribution is
the critical issue because it reflects the coupling structure of the
interactions. As such  forward-backward asymmetry $A_{\rm FB}$ in
top pair production is one of the most interesting quantities.
Sometimes $A_{\rm FB}$ is also called the charge asymmetry when CP
conservation in top sector is assumed.  Tevatron has already
observed some experimental and theoretical inconsistency in $A_{\rm
FB}$ measurements
\cite{CDFAfbnote,D0Afbnote,Abazov:2008xq,Aaltonen:2008hc,
Aaltonen:2011kc,Kuhn:1998jr,
Kuhn:1998kw,Antunano:2007da,Bernreuther:2010ny,Almeida:2008ug,
Ahrens:2010zv}. It stirred up immediately many investigations in the
BSM \cite{Frampton:2009rk, Shu:2009xf, Chivukula:2010fk,
Jung:2009jz, Cheung:2009ch, Cao:2010zb, Djouadi:2009nb, Jung:2009pi,
Cao:2009uz, Barger:2010mw, Arhrib:2009hu, Xiao:2010hm, Bauer:2010iq,
Xiao:2010ph, Dorsner:2009mq, Chen:2010hm, Cheung:2011qa}. However,
so far the precision of $A_{\rm FB}$ is limited by the small sample
collected at the Tevatron and it is hard to make a clear judgement.
In order to confirm/exclude the inconsistence, it is natural to
expect that top quark $A_{\rm FB}$ will be measured with higher
precision at the LHC, which is the top factory. If the top quark
$A_{\rm FB}$ inconsistence with the SM prediction can be confirmed
at the LHC, it will be  a sign of the BSM.

However LHC is a forward-backward symmetric proton-proton collider,
so there is no straightforward definition of $A_{\rm FB}$ as that at
the Tevatron which is a forward-backward asymmetric
proton-anti-proton collider. New observable that can reveal the
top-antitop forward-backward asymmetry, which is generated at
partonic level for example $q\bar q\rightarrow t\bar t$, is needed
at the LHC. There are some existing observables in literatures that
can fulfill this need \cite{Langacker:1984dc, Dittmar:1996my,
Petriello:2008zr, Li:2009xh, Diener:2009ee, Diener:2010sy,
Wang:2010du,Wang:2010tg,Dvergsnes:2004tw,Kuhn:1998jr,
Kuhn:1998kw,Antunano:2007da, Ferrario:2008wm}. However, each of them
poses some advantages and disadvantages. Generally speaking, the
favorite decay chain to tag the top quark pair is $t\bar{t}\to
b\bar{b}2j l \nu$, which implies that the top (or anti-top) decays
semi-leptonically in order to label the mother particle charge.
Although some techniques can be adopted, such as requiring the
invariant mass of the lepton and the neutrino should be just equal
to the $W$ mass, the undetected by-product neutrino may still cause
the non-negligible uncertainty during the event reconstruction. The
precision of forward-backward asymmetry will be limited by this
uncertainty. As such it is better not to use the momentum
information of semi-leptonically decaying top (anti-top) quark. In
this paper only hadronically decaying top (anti-top) quark momentum
information is utilized.

In order to isolate the asymmetric events from the symmetric ones
which is mainly due to the symmetric gluon-gluon fusion processes,
some kinematic region should be chosen. The requirement that the
rapidity of top is larger than a critical value $Y_{\rm C}$ can
greatly suppress the symmetric cross section. In this paper, we
define a new charge asymmetry observable in $t\bar{t}$ production at
the LHC, namely the edge charge asymmetry $A_{\rm E}$ (cf. Eq.
\ref{aedef}). In some sense $A_{\rm E}$ is an optimized version of
the central charge asymmetry $A_{\rm C}$ \cite{Kuhn:1998jr,
Kuhn:1998kw,Antunano:2007da, Ferrario:2008wm}. $A_{\rm E}$  is free
from the uncertainty of neutrino momentum reconstruction and much
larger than $A_{\rm C}$ since $A_{\rm E}$ is much less polluted by
the symmetric $gg\rightarrow t\bar t$ contributions.

In section \ref{define}, we present the definition of the edge
charge asymmetry $A_{\rm E}$. Its relation with the central charge
asymmetry $A_{\rm C}$ is discussed. In section \ref{numerical},
numerical results for $A_{\rm E}$ up to NLO QCD is calculated. In
section \ref{conclusion}, we give our conclusions and discussions.

\section{The edge charge asymmetry in top pair production at the LHC \label{define}}

As mentioned in above section, the new edge charge asymmetry $A_{\rm
E}$ satisfies: (a) utilizing single top (anti-top) kinematical
information rather than the top pair information to avoid the
uncertainty in neutrino reconstruction; (b) suppressing symmetric
$gg\to t\bar{t}$ background events as much as possible. The edge
charge asymmetry $A_{\rm E}$ is defined as
\begin{equation}
A_{\rm E}(Y_{\rm C},Y_{\rm{max}}) \equiv \frac{\sigma_{\rm t}(Y_{\rm
C}<|Y_{\rm t}|<Y_{\rm{max}})-\sigma_{\rm{\bar t}}(Y_{\rm
C}<|Y_{\rm{\bar t}}|<Y_{\rm{max}})} {\sigma_{\rm t}(Y_{\rm
C}<|Y_{\rm t}|<Y_{\rm{max}})+\sigma_{\rm{\bar t}}(Y_{\rm
C}<|Y_{\rm{\bar t}}|<Y_{\rm{max}})}\equiv \frac{\sigma_{\rm
E}^A(Y_{\rm C},Y_{\rm{max}})}{\sigma_{\rm E}(Y_{\rm
C},Y_{\rm{max}})}
\label{aedef}
\end{equation}
where rapidity $Y_{\rm C}$ is the border between the edge and the
central regions, and $Y_{max}$ is the maximum value that the
detector can cover. An ideal detector has $Y_{max}=\infty$. $A_{\rm
E}$ is the ratio of the difference and sum of the number of $t$ and
$\bar{t}$ events that fall in the edge region of the detector. Here
$t$ and $\bar{t}$ are unnecessarily from the same quark pair.

$A_{\rm E}$ depends on the choice of $Y_{\rm C}$ and $Y_{max}$.
$Y_{max}$ is determined by the geometry of the detector and $Y_{\rm
C}$ should be taken at its optimal value to obtain the most
significant $A_{\rm E}$. We will investigate the optimal $Y_{\rm C}$
at LHC in section III.

As a comparison, the so called central charge asymmetry is defined
as \cite{Kuhn:1998jr, Kuhn:1998kw,Antunano:2007da,
Ferrario:2008wm} \begin{equation}
A_{\rm C}(Y_{\rm C}) \equiv \frac{\sigma_{\rm t}(|Y_{\rm t}|<Y_{\rm
C})-\sigma_{\rm{\bar t}}(|Y_{\rm{\bar t}}|<Y_{\rm C})} {\sigma_{\rm
t}(|Y_{\rm t}|<Y_{\rm C})+\sigma_{\rm{\bar t}}(|Y_{\rm{\bar
t}}|<Y_{\rm C})}\equiv \frac{\sigma_{\rm
C}^A(Y_{\rm C})}{\sigma_{\rm C}(Y_{\rm
C})}. \label{acdef}
\end{equation}
It can be seen that the difference between $A_{\rm E}$ and $A_{\rm
C}$ is that they are defined in different regions. As symmetric
$gg\to t\bar{t}$ events are mostly located in the central regions,
the expected value of $A_{\rm E}$ should be larger than that of the
$A_{\rm C}$. For the $t\bar{t}$ events at the LHC, in the edge
region $Y>Y_{\rm C}$, the number of $t$ events will be a bit larger
than the number of the $\bar{t}$ events. Oppositely, in the central
region $Y<Y_{\rm C}$, the number of $\bar{t}$ events will be a bit
larger than the number of the $t$ events. If we cover the total
kinematical region, the asymmetric $t$ and $\bar{t}$ events in
central and edge region will be canceled completely out.

\begin{figure}[htbp]
\centerline{\hbox{
\includegraphics[height=2.5cm,width=2.5cm]
{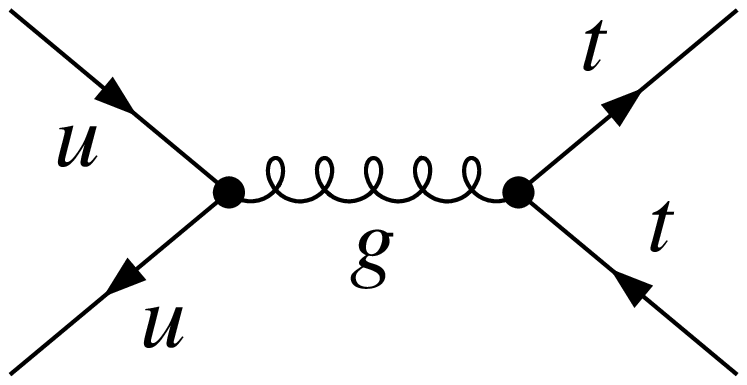}
\includegraphics[height=2.5cm,width=2.5cm]
{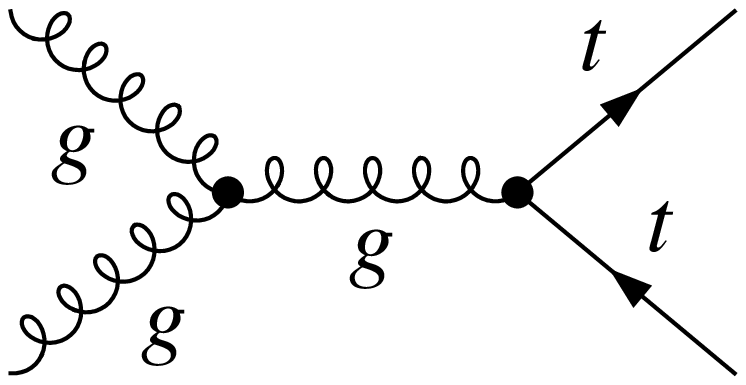}
\includegraphics[height=2.5cm,width=2.5cm]
{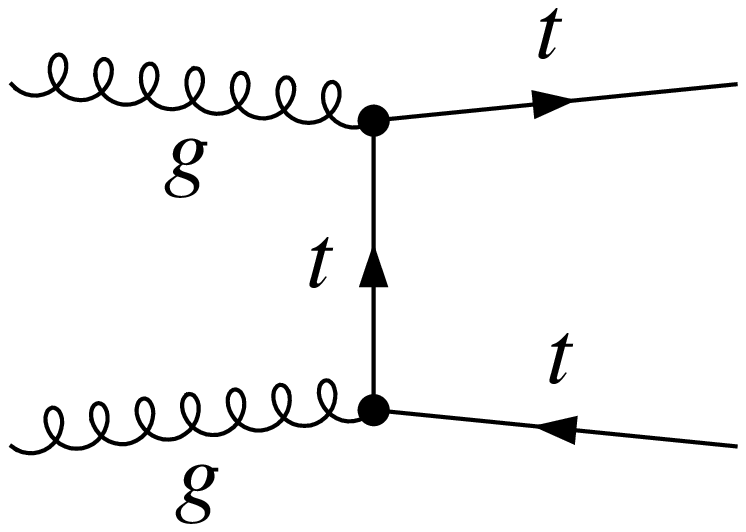} }} \caption{\label{tree} Typical Feynman
diagrams for $t\bar t$ pair production at LHC at $O(\alpha_s^2)$. }
\end{figure}

In the SM, the leading order QCD $t\bar{t}$ producing cross section
is symmetric, and the asymmetric $t\bar{t}$ cross section arise from
the next-to-leading order (NLO) QCD at the partonic level, which has
already been well studied in many literatures. In the calculation of
$A_{\rm E}$, the asymmetric cross section in the numerator is up to
NLO QCD, the total cross section in the denominator is taken as the
LO QCD symmetric cross section Fig.\ref{tree}, so as $A_{\rm E}$ is
up to $O(\alpha_s)$. Other higher order correction such as
electro-weak contribution is ignored here. The calculation are
carried out with the help of FeynCalc, FormCalc, and QCDLoop
\cite{Mertig:1990an, Hahn:1998yk, Ellis:2007qk}.

\begin{figure}[htbp]
\centerline{\hbox{
\includegraphics[height=3.0cm,width=3.0cm]
{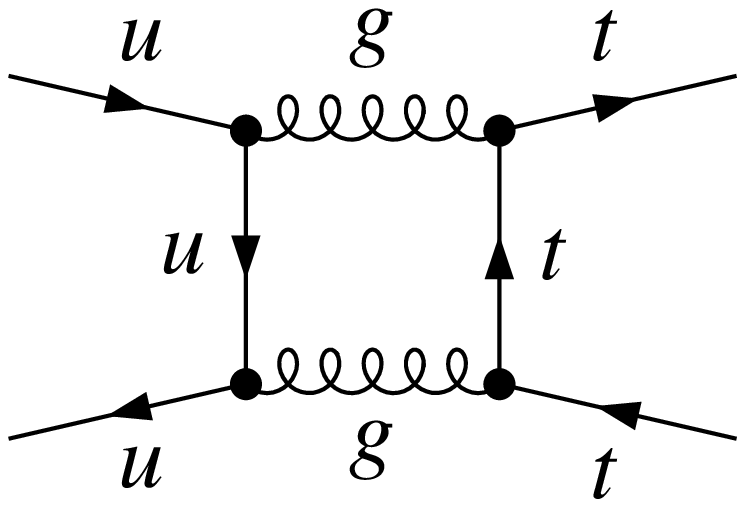}
\includegraphics[height=3.0cm,width=3.0cm]
{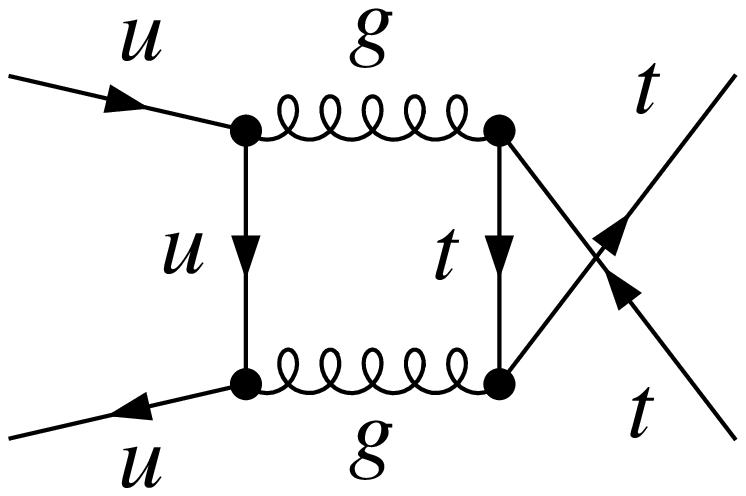} }} \caption{\label{box} Typical NLO
 virtual Feynman diagrams which contribute to asymmetric cross section. }
\end{figure}

\begin{figure}[htbp]
\centerline{\hbox{
\includegraphics[height=2.5cm,width=2.5cm]
{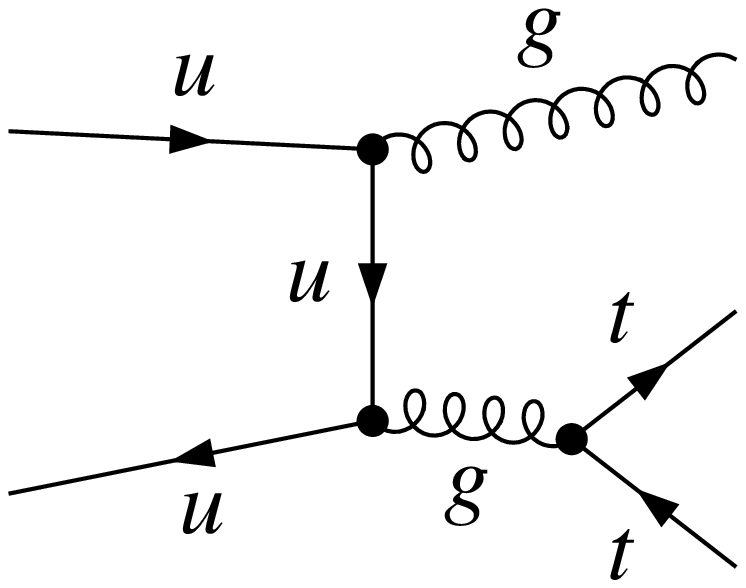}
\includegraphics[height=2.5cm,width=2.5cm]
{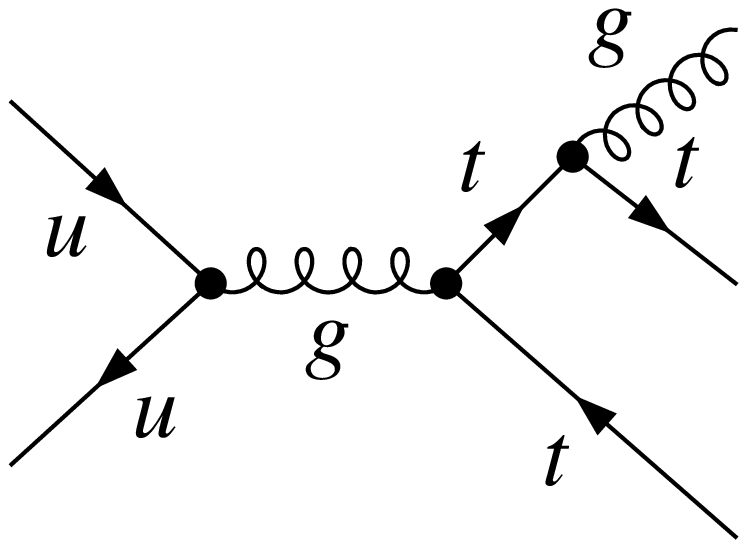} } } \caption{\label{uuttg} Typical real
gluon emission Feynman diagrams which contribute to asymmetric cross
section. }
\end{figure}

\begin{figure}[htbp]
\centerline{\hbox{
\includegraphics[height=2.5cm,width=2.5cm]
{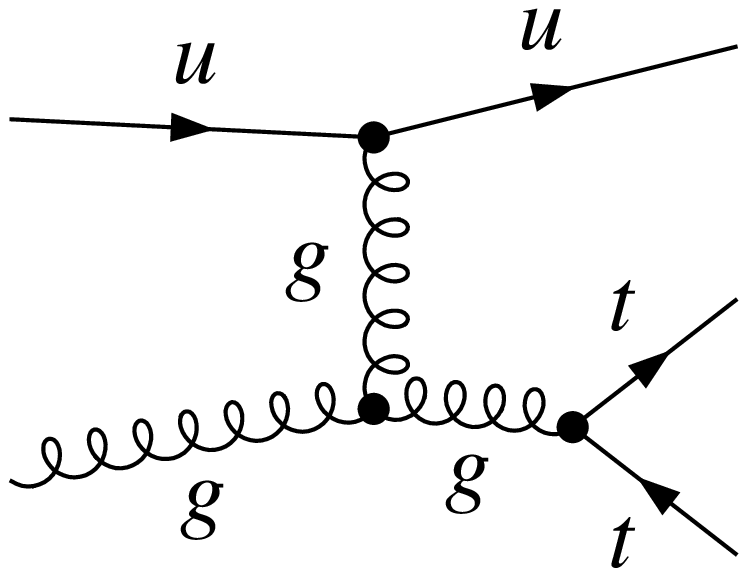}
\includegraphics[height=2.5cm,width=2.5cm]
{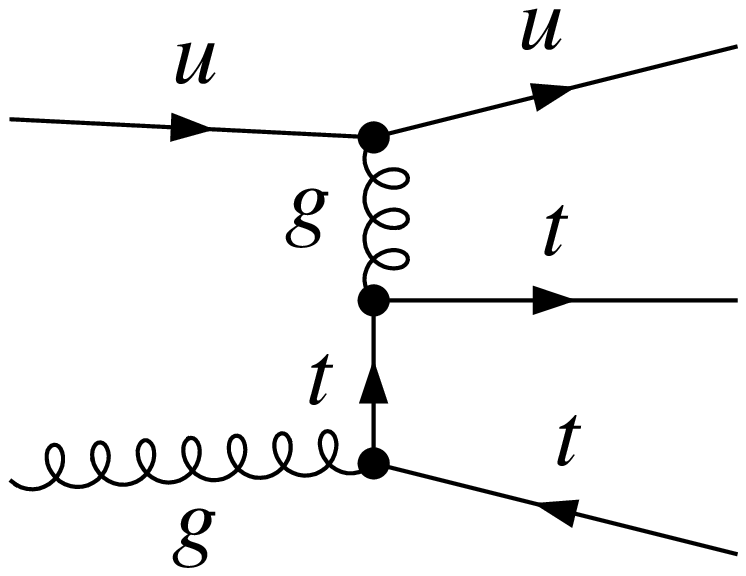}
\includegraphics[height=2.5cm,width=2.5cm]
{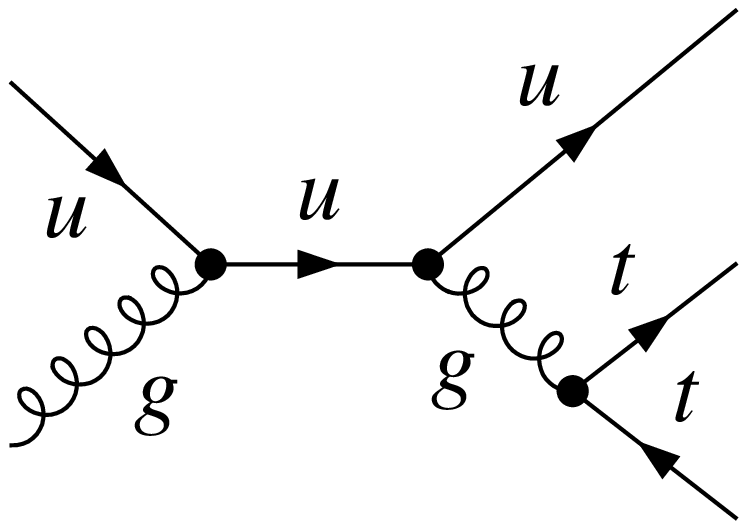} }} \centerline{(a)} \centerline{\hbox{
\includegraphics[height=2.5cm,width=2.5cm]
{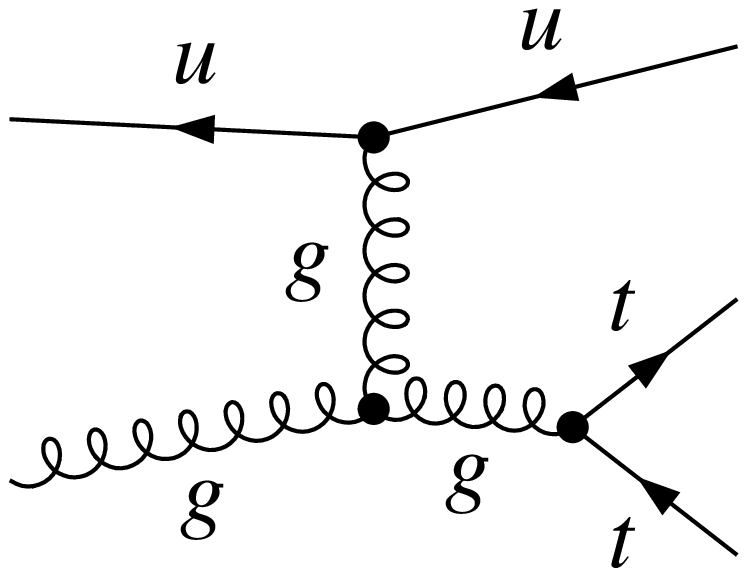}
\includegraphics[height=2.5cm,width=2.5cm]
{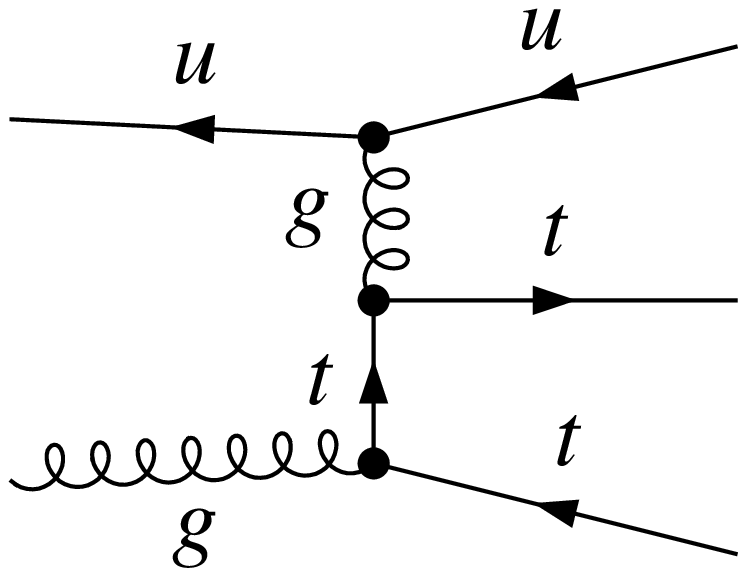}
\includegraphics[height=2.5cm,width=2.5cm]
{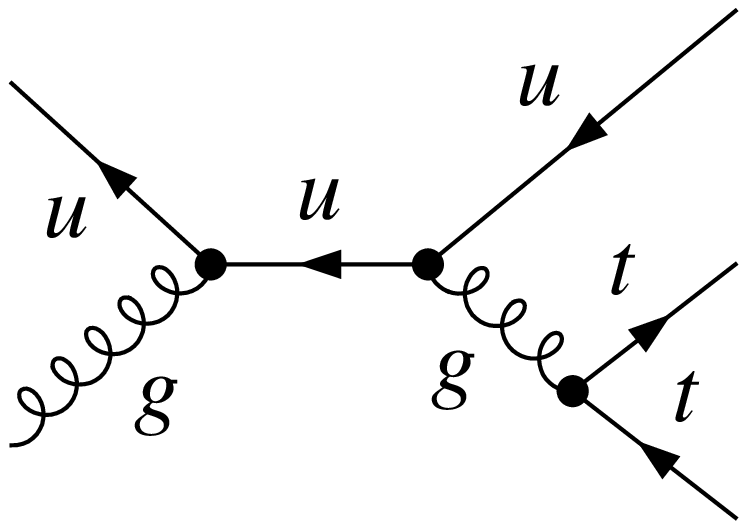} }} \centerline{(b)} \caption{\label{ugttu}
Typical Feynman diagrams of $u g \to u t\bar{t}$ (a), and $\bar{u} g
\to \bar{u} t\bar{t}$ (b).}
\end{figure}

Up to NLO QCD, $\sigma^A_{\rm E}$ gets contributions from: (1) the
interference among virtual box in Fig.~\ref{box} and the leading
diagrams for the process $q\bar q \rightarrow t \bar t$ in
Fig.~\ref{tree}; (2) the interference among initial and final gluon
radiation diagrams of $q\bar q \rightarrow t \bar t g$ in
Fig.~\ref{uuttg}; and (3) contributions from diagrams of  $q g\to
t\bar{t}q$ and $\bar{q} g\to t\bar{t}\bar{q}$ in Fig.~\ref{ugttu}.
Pay attention that the above mentioned processes does not contain
ultra-violet divergence so renormalization is unnecessary in the
calculation. Moreover, $\sigma(|Y_t| < Y_{\rm C})$ and
$\sigma(|Y_{\bar t}| < Y_{\rm C})$ contain collinear divergence
respectively, but the divergences cancel completely out when
calculating the asymmetric cross section. Soft divergences are
contained in the former (1)virtual box and (2)real radiation
contributions, but are canceled after adding the two. Technically a
soft cut $\delta^s$ is introduced after the soft divergence
cancelation\cite{Harris:2001sx}. The final results are
$\delta^s$-independent, which is carefully checked in our
calculation.

\section{Numerical Results \label{numerical}}

In the numerical calculations, the SM parameters are chosen to be
$m_t=170.9\mbox{GeV}$ and $\alpha_S(m_Z)=0.118$. We choose cteq6l
for leading order calculation and cteq6m for NLO calculations. The
scales are chosen as $\mu_r=\mu_f=m_t$.

\begin{figure}[htbp]

\begin{center}
\includegraphics[width=0.40\textwidth]
{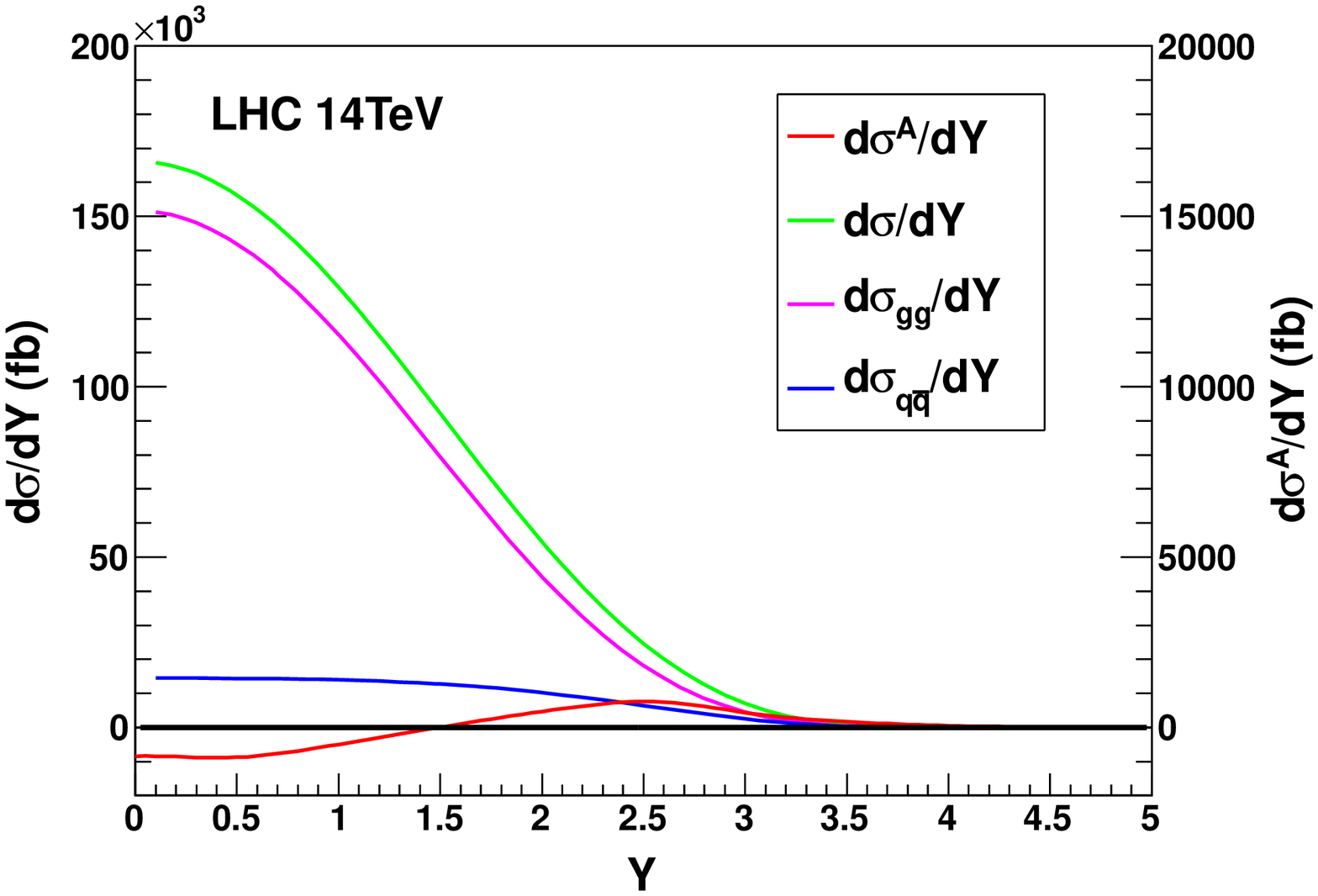}
\includegraphics[width=0.40\textwidth]
{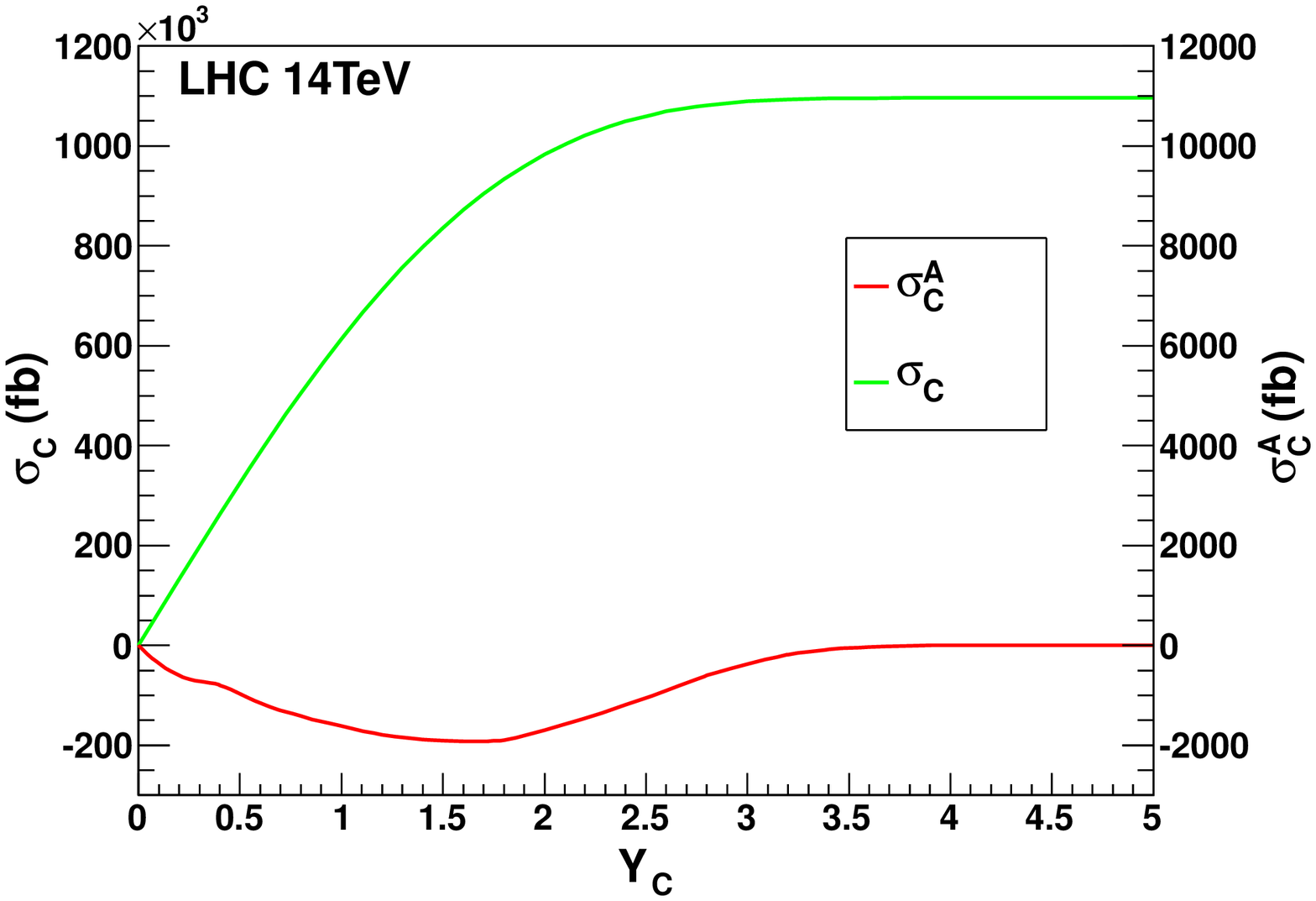}
\end{center}
\begin{center}
\includegraphics[width=0.40\textwidth]
{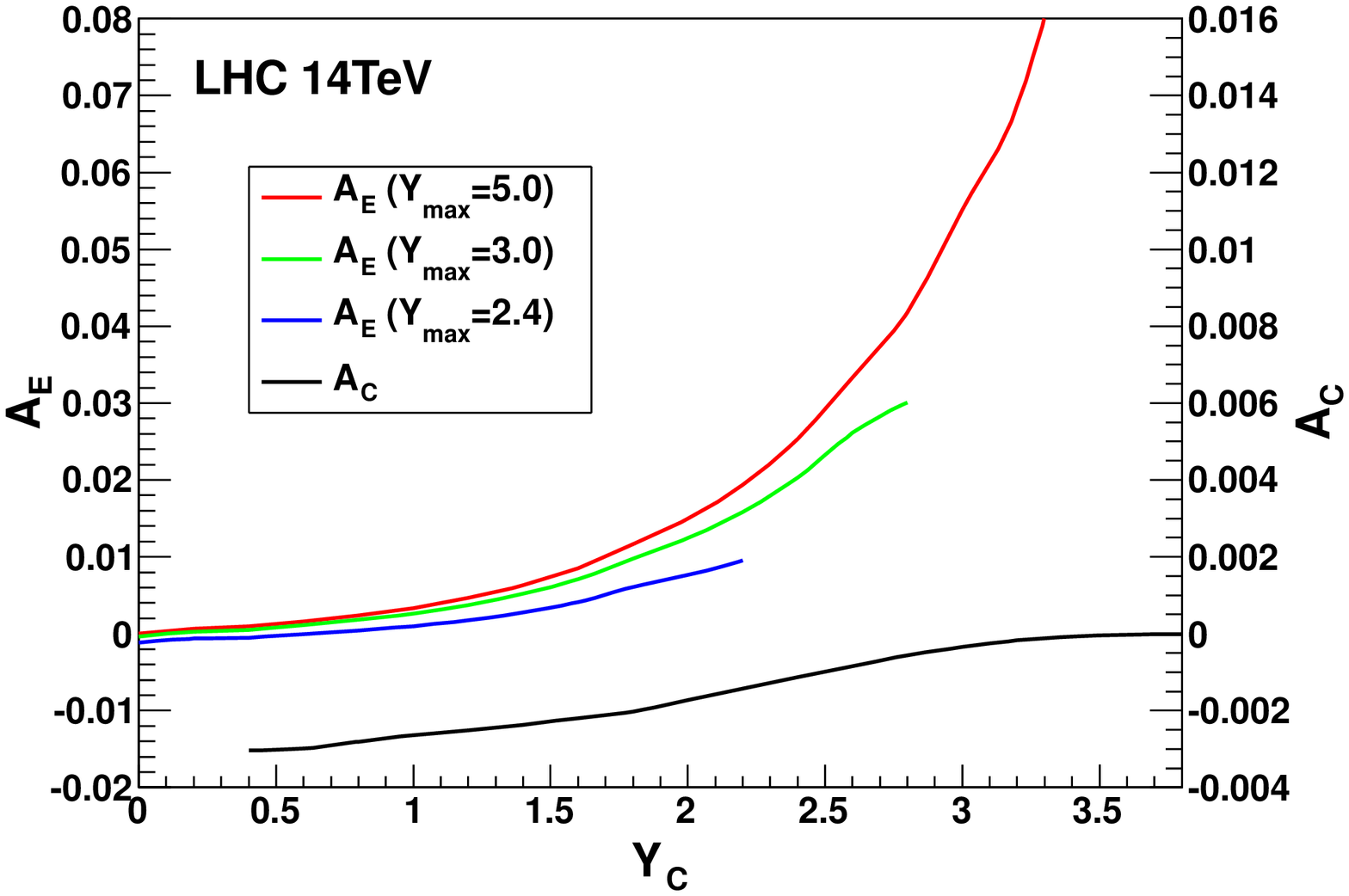}
\includegraphics[width=0.40\textwidth]
{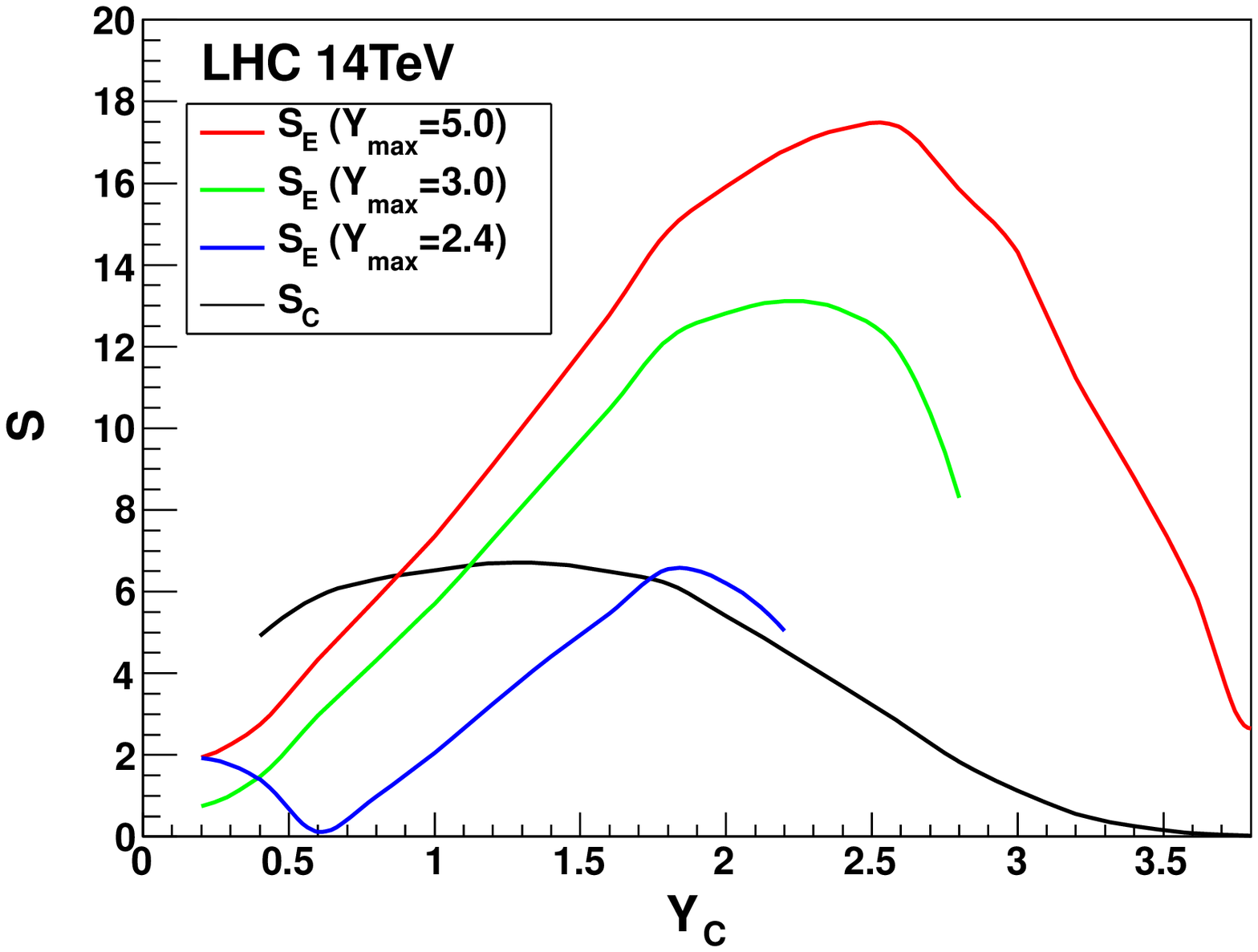}
\end{center}

\caption{\label{LHC14TeV} Left-up plot: symmetric and asymmetric
differential cross sections as a function of rapidity of top quark,
and the separate contributions to symmetric cross sections from
$q\bar q$ and $gg$ fusions are also shown. Right-up plot: symmetric
and asymmetric total cross sections in $A_{\rm C}$ (cf. Eq.
\ref{acdef}) as a function of $Y_{\rm C}$. Left(Right)-down plot:
$A_{\rm E}$ (significance $S_{\rm E}$ with integrated luminosity
$10^{-1}$fb, see text) as a function of $Y_{\rm C}$ for several
$Y_{max}=2.4,3.0, 5.0$ respectively, $A_{\rm C}$ ($S_{\rm C}$) is
also shown here. Here $\sqrt{s}$ at LHC is chosen as $14\mbox{TeV}$
and $d\sigma^A/dY$, $\sigma^A_{\rm C}$ and $A_{\rm C}$ are labeled
in the right side of the plots due to their small values. }
\end{figure}

Fig. \ref{LHC14TeV} shows the numerical estimations for the LHC with
$\sqrt{s}=14 \mbox{TeV}$. The left-up plot is the symmetric and
asymmetric differential distribution as a function of the rapidity
of $t$ or $\bar{t}$. Notice that they are labeled in different
scales. Also shown are the separate contributions to symmetric cross
section from $q\bar q$ and $gg$ fusion processes. As can be seen the
symmetric events dominantly come from the gg fusion processes and
lie mainly in the small Y region. On the contrary the asymmetric
cross section changes sign around $Y=1.6$. Namely in the central
region, the number of $\bar{t}$ events is larger than that of the
$t$ events. Oppositely, in the edge region, the number of $t$ events
is larger than that of the $\bar{t}$ events. This feature can be
easily understood as following. The asymmetric cross section will be
completely canceled out after integrating over the whole $Y>0$
region. Therefore there should be a turning point where asymmetric
cross section turns into the opposite sign. These behaviors can also
be extracted from the right-up plot, which show the symmetric and
asymmetric cross sections (cf. Eq. \ref{acdef}) as a function of
$Y_{\rm C}$. As a cross check, our result of the total leading order
$t\bar{t}$ cross section is $548\mbox{pb}$, which is consistent with
the LO QCD prediction $583^{+165}_{-120}$ in Ref.
\cite{Cacciari:2008zb}. In the left(right)-down plot in Fig.
\ref{LHC14TeV} we shown $A_{\rm E}$ (significance $S_{\rm E}$) as a
function of $Y_{\rm C}$ for several $Y_{max}=2.4,3.0, 5.0$
respectively. Significance is defined as
$S=|N^A|/\sqrt{N}=\sqrt{\cal L} |A| \sqrt{\sigma}$. Here $N^A$ ($N$)
is the number of asymmetric (symmetric) events, and the integrated
luminosity is chosen to be ${\cal L}=10\mbox{fb}^{-1}$ as an
example. In the numerical estimations we take three $Y_{max}$ values
according to the coverage of the real detectors. $Y_{max}=2.4$ is a
conservative choice and $Y_{max}=5.0$ is an optimal one. $A_{\rm C}$
($S_C$) is also shown here. From the plots, we can see clearly the
central asymmetry $A_{\rm C}$ is negative and the edge charge
asymmetry $A_{\rm E}$ is positive. Moreover $A_{\rm E}$ is much
larger than that of $A_{\rm C}$. From curves $A_{\rm E}$ is usually
several percentages while $A_{\rm C}$ is only $O(0.1)$ percentage.
Significance is also a measure to determine the optimal choice of
$Y_{\rm C}$. The maximal significance for $A_{\rm C}$ and $A_{\rm
E}$ with $Y_{max}=2.4$ is almost the same. This is not strange
because for $A_{\rm E}$ the event numbers for both symmetric and
asymmetric are reduced greatly. Therefore the precision to measure
$A_{\rm C}$ and $A_{\rm E}$ is similar. For the bigger rapidity
coverage, the significance for $A_{\rm E}$ is much larger than that
of $A_{\rm C}$ for the optimal $Y_{\rm C}$. Based on the numerical
studies, we can conclude that the detection for larger rapidity top
quark is essential to measure $A_{\rm E}$ significantly.

\begin{figure}[htbp]

\begin{center}
\includegraphics[width=0.40\textwidth]
{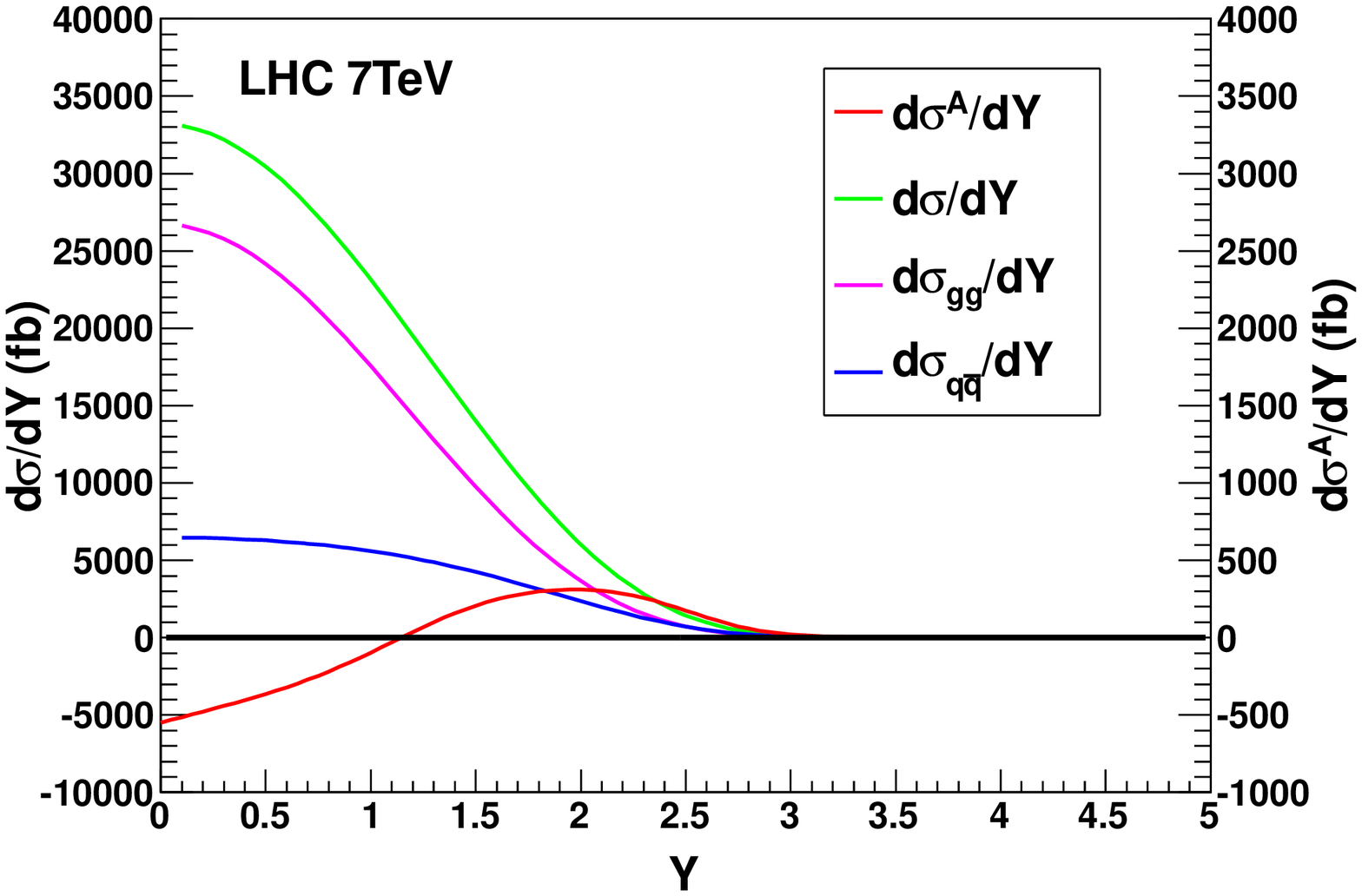}
\includegraphics[width=0.40\textwidth]
{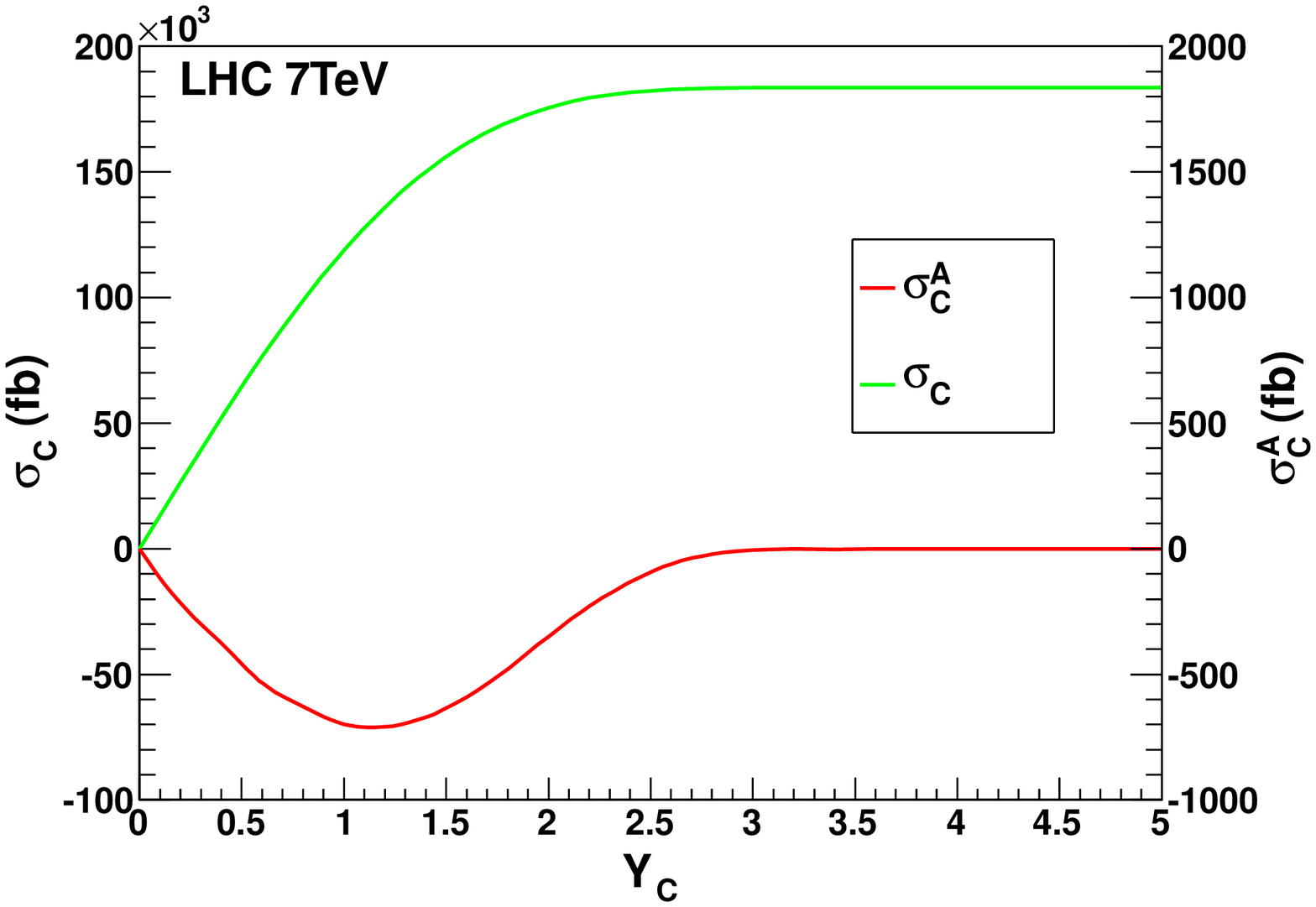}
\end{center}
\begin{center}
\includegraphics[width=0.40\textwidth]
{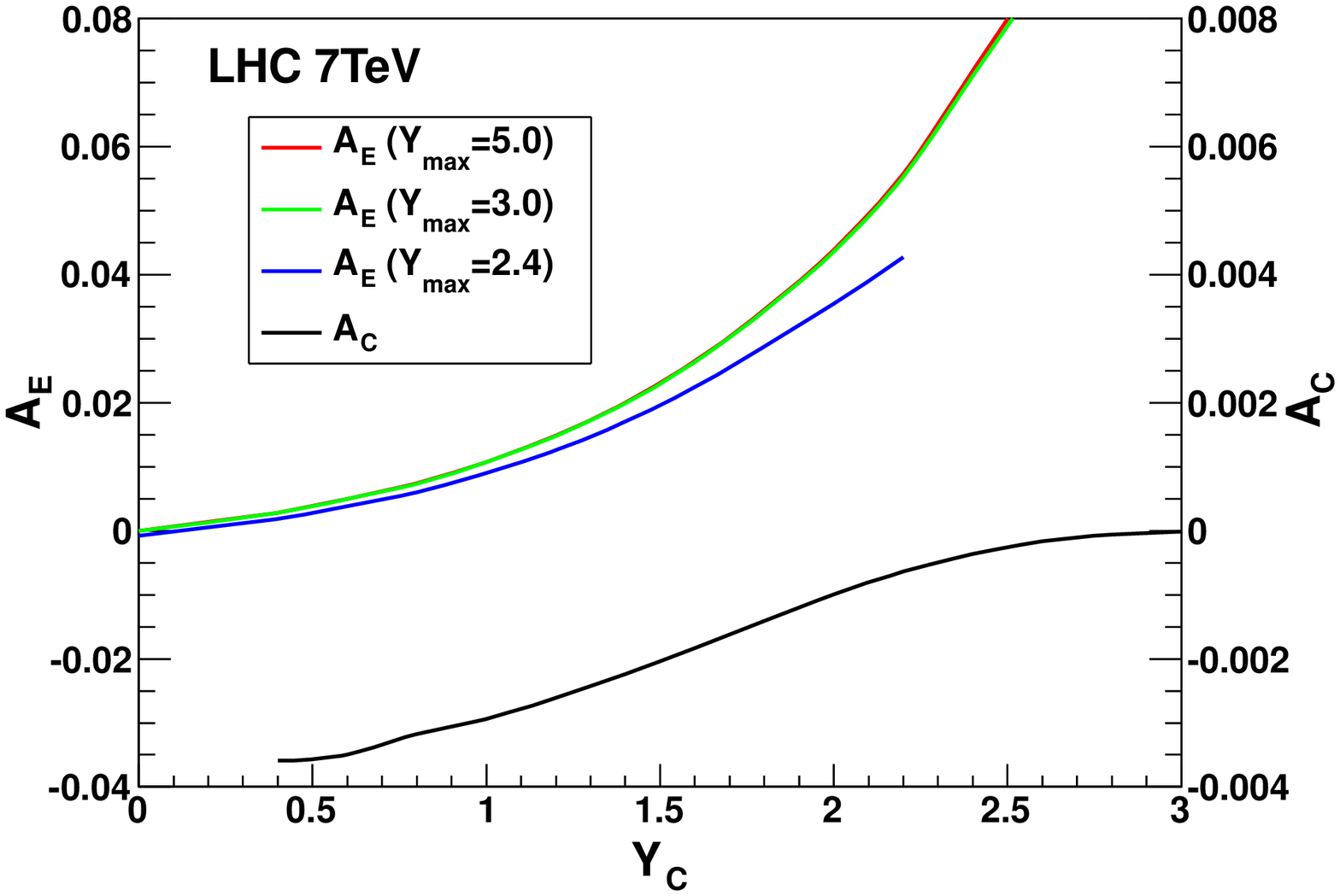}
\includegraphics[width=0.40\textwidth]
{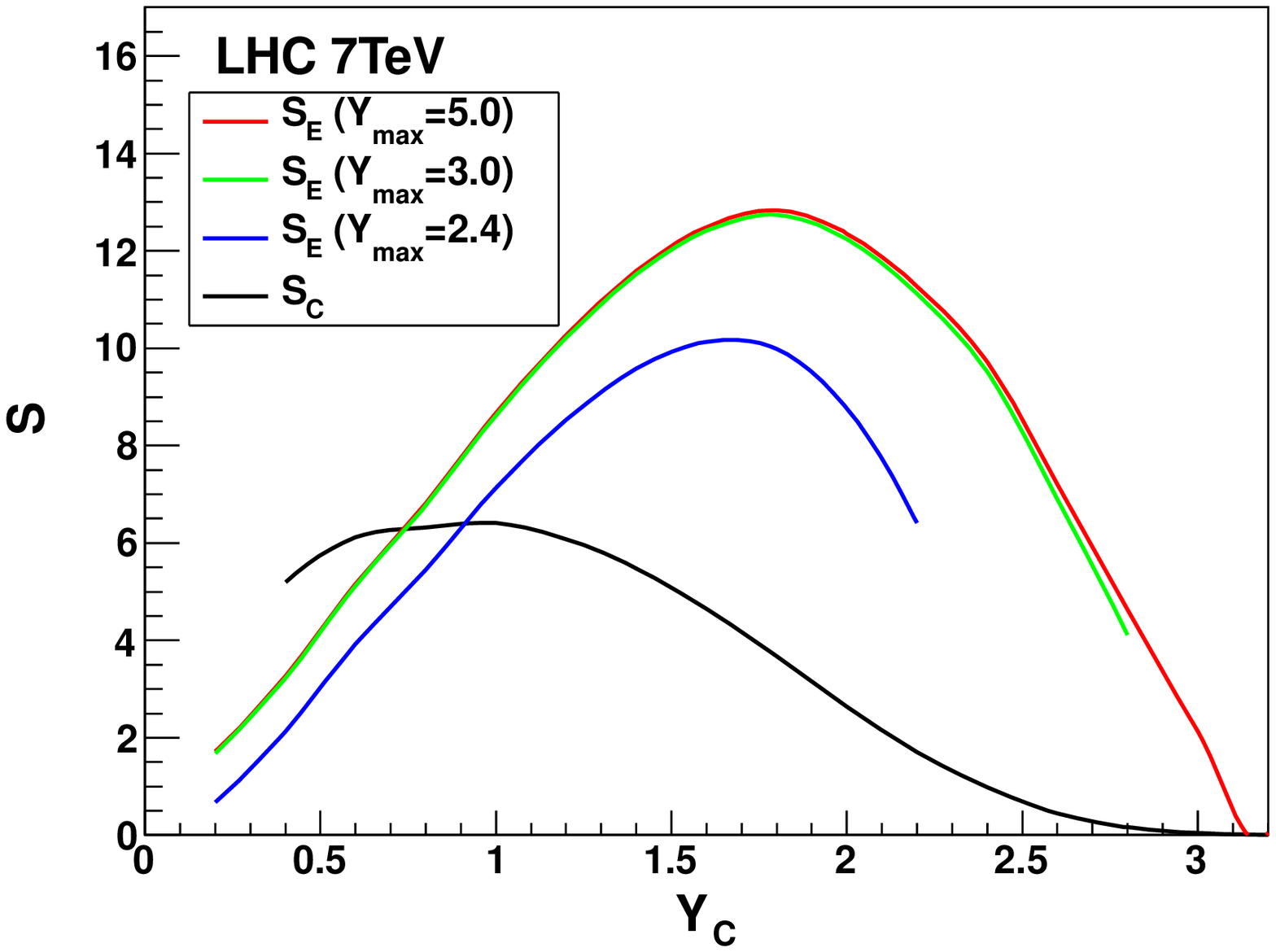}
\end{center}

\caption{\label{LHC7TeV}Same as Fig.\ref{LHC14TeV} except
for $\sqrt{s}=7\mbox{TeV}$.}
\end{figure}

Fig. \ref{LHC7TeV} shows the same distributions as those in Fig.
\ref{LHC14TeV} except for $\sqrt{s}=7 \mbox{TeV}$. Due to the lower
energy, the produced top pair events have smaller longitudinal
boosts($Y<3$). Thus curves with $Y_{max}=3.0$ and $5.0$ have small
difference. The values of the asymmetries are larger than those of
the $14\mbox{TeV}$. They are mainly caused by two effects. First, at
the parton level, a lower energy $\hat{s}$ can generate higher
asymmetry. The parton level asymmetry distribution with $\hat{s}$
can be found in ref.\cite{Kuhn:1998kw}. This can be kept at the
hadron level after the convolution of parton distribution function.
Second, the portion of the symmetric $gg\to t\bar{t}$ process become
smaller for a lower $s$. Thus the value of the charge asymmetry can
be larger with a lower $s$ than that with a higher $s$ at the LHC.

From the figures we can also see that the significance of $A_{\rm
E}$ at 7TeV is larger than that of $A_{\rm E}$ at 14TeV in the case
$Y_{max}=2.4$. The reason is that for the higher energy LHC, the top
quarks tend to be highly boosted, which shifts the distribution of
$d\sigma^A/d Y$ to the higher rapidity. After imposing $Y_{max}$
cut, the positive asymmetric cross section in the high rapidity
region losts much. Thus with the same integrated luminosity the
lower energy LHC has certain advantage to measure the top quark edge
charge asymmetry in low $Y_{max}$ case.

\section{Conclusions and discussions \label{conclusion}}

In this paper, we propose a new observable namely edge charge
asymmetry $A_{\rm E}$ in top pair production at the LHC. $A_{\rm E}$
has two advantages:  (1) free from the uncertainty arising from the
missing neutrino in the $t\bar{t}$ event reconstruction because in
the definition only single hadronically decaying top (or anti-top)
kinematical information is needed; (2) suppressing greatly the
symmetric $t\bar{t}$ events mainly due to the gluon-gluon fusion
process. Our numerical estimation showed that $A_{\rm E}$ is much
larger than that of central charge asymmetry $A_{\rm C}$
\cite{Kuhn:1998jr, Kuhn:1998kw,Antunano:2007da, Ferrario:2008wm}.
Moreover the significance to measure the $A_{\rm E}$ is usually
greater than that of $A_{\rm C}$, provided that the capacity to
identify high rapidity top quark is efficient.

\section*{Acknowledgment}
This work was supported in part by the Natural Sciences Foundation
of China (No 11075003).

\end{document}